\documentclass[12pt]{article}
\usepackage{amsmath,amsthm, amssymb}
\usepackage{graphicx}
\usepackage{leqno}
\usepackage{hyperref}
\numberwithin{equation}{section}
\begin{document}
\title{\vskip-40pt Quantum Measurements and Conservation Laws}
\author{Edward J. Gillis\footnote{email: gillise@provide.net}}

\maketitle

\begin{abstract} 

\noindent 

When a measurement is made on a system that is not in an eigenstate of the measured observable, it is often assumed that some conservation law has been violated. But this assumption results from ignoring the entanglement between the measured system and the preparation apparatus. The preparation of a system in any particular state necessarily involves interaction between the apparatus and the system, and this interaction generates entanglement. The amount of entanglement is very small, but the apparatus typically has an  enormous number of degrees of freedom; this implies that a very tiny difference between the apparatus states that are correlated with the orthogonal states of the measured system can be sufficient to account for the perceived deviation from strict conservation of the quantity in question. After reviewing entanglement relations for a typical measurement situation with a macroscopic preparation apparatus I identify microscopic processes involving analogous entanglement relations in which measurements on both entangled systems demonstrate exact conservation of the relevant quantities. The close analogy strongly suggests that conservation laws hold exactly in all measurement processes.

\end{abstract}

\section{Introduction}
\label{intro}

Measurements of quantum systems typically involve the preparation of an elementary or other microscopic target system in a particular state and a measurement apparatus designed to project the system into one of a range of several possible orthogonal states, which might or might not include the state in which the target system was prepared. Interpretive problems arise when the prepared state is \textit{not} one of the possible measurement states. One of these problems concerns what happens to putatively conserved quantities such as energy, momentum and angular momentum. To properly assess what happens to these quantities in measurement processes we need to take account of how these quantities are exchanged during the interactions that occur in both the preparation and measurement phases of the process.

It is also critical that we treat all of the systems involved, preparation, target, and measurement apparatus, as \textit{quantum} systems, and to consider the interactions between these systems as capable of exchanging the relevant quantities in \textit{either} direction. The fallacy that often occurs in analyses of the conservation problem is to regard the preparation apparatus as classical and to assume that it can act on the target system without being acted on by the target system. This assumption is a rejection of the fundamental principle that action equals reaction, and so it amounts to begging the question of whether conservation laws hold exactly in measurement processes.

One of the biggest problems in determining what happens to various quantities during measurements stems from the enormous difference in scale between the system being measured and the macroscopic systems involved in the preparation and measurement of the target system. But these differences in scale are not an absolute obstacle to assessing the effect of interactions on putatively conserved quantities. Consider a simple classical analogy. When a person bounces a ball against a wall the ball is initially headed toward the wall and after contact it is moving away from the wall. The momentum of the ball has clearly changed in a measurable way, but there is no feasible way to measure the compensating momentum change in the system with which it has interacted. But nobody regards this as evidence of a violation of conservation of momentum. We recognize that the difference in scale of the interacting systems makes it essentially impossible to test the conservation law in this circumstance. We know how to arrange many other situations in which the change in momentum of each of the interacting systems can be measured with sufficient accuracy to test whether momentum is conserved. We then use a combination of experimental results and theory to assess the principle.

Although they are often overlooked there are situations in quantum theory in which systems that can reasonably be regarded as analogous to preparation apparatuses are of small enough scale that they can be accurately measured. These will be described later. To begin it necessary to clearly define the quantities that are in question. The definition can be gleaned from the derivation of the conservation laws in quantum theory. To determine whether a quantity, $\mathbf{q}$, is conserved one must integrate the associated observable, $\mathbf{Q}$, over arbitrary states, $ |\psi\,\rangle $: 
\begin{equation}\label{1eq1}
(d\mathbf{q}/dt) \; = \; 
(d/dt)\langle\,\psi|\hat{\mathbf{Q}}|\psi\,\rangle.  \end{equation} 
If the observable commutes with the (time-independent) Hamiltonian then the quantity, $\mathbf{q}$, is conserved. 
\begin{equation}\label{1eq}
(d/dt)\langle\,\psi|\hat{\mathbf{Q}}|\psi\,\rangle \; = \; 
\frac{i}{\hbar}\langle\,\psi|\hat{\mathbf{H}}\hat{\mathbf{Q}} - \hat{\mathbf{Q}}\hat{\mathbf{H}}|\psi\,\rangle \; = \; 0.  \end{equation} 
Thus, the proper definition of a conserved quantity is given by the expression, $\langle\,\psi|\hat{\mathbf{Q}}|\psi\,\rangle$. This is a consequence of the superposition principle which is a defining feature of quantum theory. These quantities should not be seen merely as ``expectation values". They are \textit{the} quantities that quantum theory implies are conserved, provided that they meet the commutation condition. To restrict the definition of `conserved quantity' to situations in which a system that one can feasibly measure is in an exact eigenstate of the measured observable limits the application of conservation laws to a small fraction of real cases. The idea that these laws \textit{hold} only in a narrow statistical sense must be replaced by the recognition that in the kind of standard measurement scenarios described above they can only be \textit{tested} in a probabilistic manner.

Entanglement is a generic result of interaction between physical systems as demonstrated by Gemmer and Mahler\cite{Gemmer_Mahler} and by Durt\cite{Durt_a,Durt_1}. When a preparation apparatus interacts with a target system the two systems become entangled; the states of the target system become correlated with states of the apparatus. The states of the target system are orthogonal and relatively easy to distinguish, but there is very little difference between the correlated states of the apparatus. This means that the effects of a subsequent measurement on the preparation apparatus are essentially undetectable. Nevertheless, when a measurement apparatus subsequently ``collapses"\footnote{The term, `collapse', is used informally to describe particular measurement outcomes, however one interprets them. Various interpretations are discussed later.} the state of the target system the state of the apparatus is also affected. The correlation between the surviving states of the target system and preparation apparatus maintains exact conservation of the relevant quantities. This process will be outlined with a simple example in the next section. After that I will describe some cases in which exact conservation can be explicitly verified in situations that are closely analogous to macroscopic preparation/measurement scenarios.

 \section{Interaction, Entanglement, and Conservation}
 \label{sec:2}

 The interaction and resulting entanglement between a target system and a preparation apparatus can be illustrated fairly simply by considering a photon, a beam-splitter and a downstream detector in the reflected branch of the photon. When the photon interacts with the beam-splitter it is divided into transmitted and reflected branches. The reflection of one branch involves an exchange of momentum between the photon and the beam-splitter.\footnote{This example has been used in previous works by this author\cite{Gillis_3,Gillis_4}, and it has also been used by Marletto and Vedral\cite{Marletto_Vedral} in an article that makes an argument similar to the one presented here.}  Because this sort of arrangement is used in a Mach-Zehnder interferometer we know that the coherence of the photon is maintained to an extremely high degree. This implies that the uncertainty in the position of the reflecting surface must be several orders of magnitude less than the dominant wavelength of the photon. Hence the quantum state of the beam-splitter must span a range of momenta that is several orders of magnitude greater than the momentum of the reflected photon. So the difference in the states of the beam-splitter that are correlated with the reflected and transmitted branches of the photon is extremely small, but it is not zero. As mentioned above, to refuse to recognize this difference is to beg the question about momentum conservation.
 
 To reinforce this argument we can write an expression for the small amount of entanglement that is generated between the photon and the beam-splitter. The von Neumann entropy can be used as a measure of entanglement. Assume that the inner product of the two resultant beam-splitter states has a magnitude of $| 1-\epsilon |$, where $\epsilon \; \ll \; 1$, but $ \epsilon \; > \; 0 $. The von Neumann entropy formula requires that the correlation between the systems be expressed in a biorthogonal basis. To simplify the derivation we also assume that the amplitudes of the reflected and transmitted branches of the photon are equal in magnitude. The biorthogonal decomposition can be determined by taking the sum and difference of the two beam-splitter states and then normalizing. These are correlated with the sum and difference of the photon branches. This makes the entanglement calculation  fairly simple. The result can be approximated as  
$(\epsilon/2)[1-\log(\epsilon/2) ].$ 
 
 The entanglement between the photon and beam-splitter implies that a subsequent detection of the photon in the reflected branch collapses not only the state of the photon but it also collapses the state of the beam-splitter to the state that is correlated with the reflected branch. The momentum change of the photon is compensated for by the change in the momentum of the beam-splitter. Momentum is conserved. 
 
 Of course, we cannot make a sufficiently precise measurement of the beam-splitter to verify the conservation, but the situation is not essentially different in that respect  from the classical example of a ball being bounced against a wall. It is also not necessary in either the classical or quantum case to completely track the momentum change back to some beginning point. Macroscopic systems like beam-splitters and walls inevitably interact with their environment, and these interactions involve additional exchanges of momentum. In both cases we need to look for other situations in which to test conservation laws.

 The example helps make clear the problem with directly testing conservation laws in typical measurement situations. The target system is in a superposition of clearly distinguishable, orthogonal states; the particular decomposition is defined by the measurement arrangement. The conserved quantity, $\langle\,\psi|\hat{\mathbf{Q}}|\psi\,\rangle$, is given by integrating the eigenvalues of the quantity over all of the superposed states, but the selection of one of the states along with its eigenvalue changes the value of the conserved quantity exhibited by the target system. Because the difference in the correlated states of the preparation apparatus is so tiny there is no way to check whether there is a compensating change in the quantity.\footnote{In some cases the change can be attributed to the measurement apparatus. These will be discussed in Section 3.}
 
 It is the transition during measurement of the target system from a superposition of orthogonal states to a particular state that defines the issue regarding exact conservation. In order to test whether momentum is conserved exactly we would like to find situations in which two comparably sized quantum systems are in a correlated state analogous to the situation with a preparation apparatus and a target system, but where the correlated pairs of states of both systems are readily measurable. In other words, we need to identify cases in which the  biorthogonal decomposition of the entangled systems naturally picks out the states of \textit{both} systems that are easiest to measure.

 Fortunately, nature has provided us with such situations in cases involving decay processes. Consider three examples; (1) an atom in an excited state emitting a photon, (2) alpha decay, (3) beta decay. In these processes the parent particle plays a role analogous to that of the preparation apparatus in a typical quantum measurement situation. In the initial stages the total system is in an entangled state; the parent and daughter particles are in a superposition of correlated momentum states. 
 
 When an excited atom emits a photon the total system remains in the entangled state until the photon is detected. The detection collapses the superposition of momentum states of the photon to a single well defined value. The emitting atom undergoes a recoil that exactly compensates for the detected momentum state. The total momentum of the system is the same before and after the decay and detection. Energy and angular momentum are also conserved in the process. 
 
 In alpha decay a heavy nucleus emits an alpha particle (helium nucleus). The total system, initially in a superposition of states, collapses after detection of the emitted particle into a single correlated state with the same total momentum as the initial state.

 Beta decay offers a particularly instructive example of exact conservation. Initial observations of beta decay showed the emission of protons and electrons from nuclei in which the supposedly conserved quantities of all the remnant particles did not sum to zero. The insistence on exact conservation of the relevant quantities led to the prediction of the neutrino which was subsequently discovered, and found to validate the conservation laws. 
 
 To summarize, in situations in which possible changes in values of putatively conserved quantities can be accurately measured we find that these quantities are conserved exactly, whether or not the initial states of the systems involved are in eigenstates of those quantities. In cases where the interacting systems are of vastly different size it is just not feasible to measure changes in the total value of a particular quantity after a process has been completed. But this is true in \textit{both classical and quantum} measurement situations. The reason is that it is simply not feasible to track all of the interactions of large systems with their surroundings. But in neither case should we take this lack of ability to track every interaction as evidence of a violation of exact conservation laws.

 \section{Some Additional Considerations}
 \label{sec:3}

	As stated at the beginning the principal reason for questioning whether conservation laws hold exactly in measurement processes is the failure to take into account the interactions between the preparation apparatus and the target system and the entanglement that is generated. This issue has been dealt with in the previous sections. However, there are some secondary concerns regarding exact conservation that should be addressed. 
	
	The first of these has to do with apparent deviations from exact conservation that result from interactions between the target system and the measurement apparatus. These have usually been seen as much less problematic than the kinds of situations described earlier. There are several related reasons for this. The interactions between the target system and the measurement apparatus clearly change the state of the apparatus. So it is quite reasonable to assume that conserved quantities have been exchanged. Furthermore, the exchange is local, brought about directly through the interaction rather than nonlocally, through pre-established entanglement relations. A simple illustration of these kinds of cases is given by a Bell type measurement scenario. Consider a singlet state with two spin-one-half particles. Suppose a measurement on the first particle yields an x-up spin state. If the spin of the second particle is also measured along the x-axis the result will be an x-down spin state, thus conserving angular momentum exactly. But if the spin of the second particle is measured along a different axis, the sum of the two spins is no longer zero, suggesting a deviation from exact conservation. These apparent violations are resolved very readily by attributing the missing angular momentum to the measurement apparatus.

	The second issue concerns what can be called the relativity of entanglement. The argument in the previous sections relies to a large extent on the entanglement between the target system and the preparation apparatus generated by their interactions. The potential issue is that whether a state is regarded as entangled or factorizable depends on the tensor decomposition of the Hilbert space, that is, on how the total system is decomposed into subsystems. This kind of dependence has been discussed by Dug\'{i}c, Jekn\'{i}c-Dug\'{i}c, and  Arsenijev\'{i}c,\cite{Dugic_1,Dugic_2}, and also by Thirring, Bertlmann, K\"{o}hler, and Narnhofer\cite{Bertlmann_2,Bertlmann_1}. Alternate decompositions are often used to facilitate calculations. The relevant decomposition in these situations is defined by the interactions that are responsible for the exchange of conserved quantities between subsystems. This is pointed out by the authors of \cite{Bertlmann_2}:
	\begin{quote} 
		``Thus it's only the interaction, which we consider to determine the density matrix, or the measurement set-up, which fixes the factorization."
	\end{quote} 
	This point re-emphasizes the central role of interactions in the questions regarding conservation. It is interactions that both generate entanglement and redistribute conserved quantities among subsystems.

	The third issue concerns the use of an informal notion of wave function collapse used in the analysis in Sections 1 and 2. The general question of whether conservation laws hold exactly or only statistically in quantum measurements presupposes that the quantum formalism refers in some way to an underlying physical reality. If one adopts an interpretation in which the formalism is just a set of calculational recipes it is not clear that the question of exact vs. statistical conservation can even be formulated. 
	
	Realistic interpretations fall roughly into three main categories: (1) many worlds-decoherence, (2) de Broglie-Bohm pilot wave theories, (3) wave function collapse proposals. In the many worlds/decoherence interpretation apparent collapse is just a by-product of the separation of the wave function into clearly distinct branches. The separation of the branches is induced by conservative interactions. This means that the conserved quantities are redistributed between the interacting subsystems, but the total amount of the quantity in each branch is unchanged (up to normalization of the branch). Use of the term, `collapse', in this context just provides a convenient way of comparing the before-and-after values of the relevant quantities. The key point, as just stated, is that conserved quantities are just redistributed - total values are not changed. 
	
	The analysis of conservation in pilot wave theories is essentially the same as that just described for many worlds/decoherence accounts. The overall wave function evolves and divides into branches in exactly the same way. The difference between the two approaches is that pilot wave theories assume that there is a system point, which is the point in configuration space that encodes the positions in physical space of all of the particles that make up the total system. This system point is present in exactly one branch at any given time, and this is the branch that is actualized. The quantities of interest are those computed from the wave function - not those describing the motion of the individual particles. Since conservation is exact (up to normalization) in every branch of the wave function the analysis is the same as that in the previous paragraph. 
	
	The term, `collapse', is interpreted literally by those who regard wave function collapse as an actual physical occurrence (or process). If one views collapse as an instantaneous occurrence in which one of the several branches is selected in conformity to the Born probability rule then there is no problem with exact conservation. The analysis is the same as that for decoherence or pilot wave accounts. But most attempts to describe it as a real occurrence propose nonlinear, stochastic modifications to the Schr\"{o}dinger equation and treat it as a process. In many of these the mechanisms introduced to generate collapse also induce small violations of energy conservation and possibly conservation of other quantities. It is necessary to emphasize, however, that these sorts of violations are independent of the issues that were discussed in the previous sections. Moreover, one such proposal has shown how to generate actual collapse while guaranteeing exact conservation of the relevant quantities.\cite{Gillis_JPA1} This is accomplished, in part, by constructing the stochastic collapse operator from conservative interaction potentials.

	This has been an attempt to deal with the questions about conserved quantities in quantum measurements in a fairly brief and compact manner. So to close this section it is worth pointing the reader to some of the other works that have addressed various issues involved in the relationship between quantum measurements and conservation laws. Arguments along some of the same lines as those presented here have been given in 
	\cite{Zeng_Sun_Shao,APR_2,APR_2_seq}. Counter arguments have been presented in \cite{Carroll_Lodman,OSW}. These issues have been discussed in connection with quantum clocks in \cite{Gisin_Q_Clocks}. Early work on some related issues includes  \cite{WAY_1,WAY_2,WAY_3}. Some related issues are also discussed in \cite{Stoica_Block}. This list is not in any way exhaustive. Additional references can be found in the cited works

\section{Summary}
\label{sec:4}

   The question of whether conservation laws hold exactly in quantum measurement processes presupposes that: (1) the quantum formalism refers in some way to an underlying physical reality; (2) all systems, both microscopic and macroscopic are \textit{quantum} systems. The first of these assumptions must be made to formulate the question of exact conservation. If the quantum formalism is considered to be just a probabilistic calculus then the question would be ill posed. The second assumption is necessary to avoid begging the question of exact conservation. To allow large systems to act on small systems without allowing small systems to act on and possibly alter large systems is a rejection of the principle that for every action there is an equal and opposite reaction. This principle is central to conservation laws. 
   
   Given this framework, we know that under unitary evolution conserved quantities are exchanged between subsystems during interactions. Measurement processes typically start with the separation of wave function components into distinct branches corresponding to different possible measurement outcomes. This separation is brought about by conservative interactions. The relevant quantities are exchanged between subsystems. So the distribution of those quantities among the component subsystems  varies from one branch to another, but the total quantities are the same in every branch (up to normalization of the branch). 
   
   The question of conservation then comes down to how one views the ``collapse" of the wave function that is associated with measurements. As described in Section 3 there are three general interpretive approaches that regard the quantum formalism as referring to an underlying physical reality. Two of these are the many worlds/decoherence (Everett/Zurek) interpretations and pilot wave (de Broglie-Bohm) theories. In these interpretations there is no actual collapse of the wave function. In Everett interpretations the appearance of collapse is simply a result of the observer of the measurement outcome branching along with the target system and measurement apparatus. In pilot wave theories the apparent selection of one particular branch is just a reflection of the fact that the system point resides in exactly one branch of the wave function. Therefore, in either of these two realistic interpretations of quantum theory conservation laws hold exactly in measurement processes. 
   
   The third type of realistic interpretation treats wave function collapse as a real physical occurrence. If one assumes that the collapse is simply an instantaneous event that picks out one of the branches in accord with the Born rule then, again, conservation laws hold exactly. However, most proponents of this approach try to incorporate a description of wave function collapse into the mathematical formalism of quantum theory by adding stochastic, nonlinear terms to the Schr\"{o}dinger equation in a way that induces collapse in accord with the Born rule. In many of these proposals the stochastic operator and stochastic process (or processes) that induce collapse also generate small violations of conservation laws (even in a statistical sense). However, there is a proposal in which the stochastic operator is constructed from conservative interaction potentials, which ensures exact conservation in individual cases.\cite{Gillis_JPA1} Therefore, the view that wave function collapse is a real physical occurrence is consistent with exact conservation in measurement processes. 
   
   Experimental evidence for exact conservation is provided by examination of decay processes as described in Section 2. In those processes a ``parent" system such as an excited atom, radioactive nucleus, or an unstable particle like a neutron emits a ``daughter" particle. The parent-daughter system is in an entangled state analogous to the state of a preparation apparatus-target system in a typical macroscopic measurement situation. The only difference is that in the decay processes the correlated states of the parent system are readily distinguishable and easily measurable. The numerous experiments that have been done are consistent with exact conservation. The combination of theoretical and experimental considerations strongly support the claim that conservation laws hold exactly in measurement processes.

\section*{Acknowledgments} 
I would like to thank Ovidiu Cristinel Stoica and Nicolas Gisin for bringing additional references and viewpoints to my attention.

\end{document}